\documentclass[floatfix,reprint,amsmath,amssymb,aps,pra]{revtex4-2}
\usepackage{times,mathptmx}
\usepackage{graphicx}
\usepackage[english]{babel}
\usepackage{epstopdf}
\usepackage{longtable}
\usepackage{xcolor}

\begin{document}

\title{Controlling exceptional points of layered non-Hermitian structures with magnetic field}
\author{Milena~Ramanovich$^{1}$}
\author{Denis~V.~Novitsky$^{1,2}$}

\affiliation{$^1$B.~I.~Stepanov Institute of Physics, National Academy of Sciences of Belarus, Nezavisimosti Avenue 68, 220072 Minsk, Belarus \\
$^2$Department of Laser Physics and Spectroscopy, Physics Faculty, Belarusian State University, Nezavisimosti Avenue 4, 220030 Minsk, Belarus}

\date{\today}

\begin{abstract}
Exceptional points are the characteristic features of non-Hermitian systems providing numerous applications in optics, acoustics, electronics, etc. We show that exceptional points can be controlled with the external magnetic field by considering the archetypal three-layered $\mathcal{PT}$-symmetric photonic structure including a magneto-optic material between the loss and gain layers. Application of the magnetic field makes the system anisotropic with the polarization-sensitive reflection and transmission properties. As a result, the magnetic field can stimulate $\mathcal{PT}$-symmetry breaking and mode exchange at realistic values of system parameters. Moreover, we suggest that the proposed scheme can be also used for magnetic-perturbation sensing with an easily adaptable level of desired field strength. Our results contain a valuable evidence for the usefulness of anisotropic (including magnetic) materials for tunable non-Hermitian photonics.
\end{abstract}

\maketitle

\section{Introduction}

Exceptional point (EP) is a peculiar singularity in the parameter space of physical systems described with non-Hermitian operators (matrices), such as Hamiltonian or scattering matrix. At the EP, both eigenvalues and eigenvectors of the system's operator coalesce making its modal composition incomplete. As a result, the behavior of system becomes nontrivial close to the EP. Although the EPs and non-Hermitian dynamics in general were first studied in quantum mechanics and are generally applicable to physical systems of arbitrary nature \cite{Ashida2020, Xiao2025}, these features proved to be highly relevant for modern photonics providing novel approaches to optical structure composition and promising unexpected applications \cite{Miri2019, Li2023, Novitsky2024}. In particular, non-Hermiticity is often introduced through optical loss and gain media, as dramatically exemplified in $\mathcal{PT}$-symmetric structures \cite{Zyablovsky2014, Feng2017, Ganainy2018, Ozdemir2019, Krasnok2021, Kim2026}.

As to applications of EPs, one can readily mention enhanced sensing due to sublinear (root-like) splitting of eigenvalues as a function of perturbation strength. This is perhaps the most discussed and promising application of EPs \cite{Wiersig2020, Wang2025, Wu2025}. Although the straightforward EP-based sensors proposed in the pioneering publications \cite{Chen2017, Hodaei2017} seem to be non-advantageous due to quantum-noise enhancement and, hence, reduced signal-to-noise ratio \cite{Duggan2022, Ding2023, Loughlin2024, Zheng2025, Almanakly2026}, there is a number of advanced approaches to circumvent the problem, such as detuning from the EP \cite{Xiao2024, Wang2026a, Landers2026} or using nonreciprocity \cite{Lau2018}, nonclassical states of light \cite{Zhang2019a, Wang2026b}, and unconventional non-Hermitian matrices, such as the Jacobian matrix \cite{Cognee2026}. Since non-Hermitian photonic systems often contain nonstandard distributions of loss and gain, they form the basis for novel effects of controlled lasing \cite{Feng2014, Hodaei2014, Qi2019} and perfect absorption \cite{Chong2010, Baranov2017}, including the combined laser-absorber regime in $\mathcal{PT}$-symmetric structures \cite{Longhi2010, Novitsky2019}. EPs are deeply connected to nontrivial topological properties of non-Hermitian systems \cite{Wang2021, Parto2021} resulting in such effects as the topological mode switching \cite{Dembowski2001, Doppler2016, Zhang2024, Li2025, Hong2025, Park2026} and interplay with the non-Hermitian skin effect \cite{Wang2025a, Zhang2025}. Finally, we mention the effects of asymmetric light propagation and generation in non-Hermitian photonic structures, which are deeply connected to EPs \cite{Ge2012, Novitsky2018}.

For possible applications, it is of vital importance to control EPs on demand, so that one could change the position of EPs or tune the system closer or farther from an EP with an external influence. Such tunability will allow us to manipulate EPs in real time without the necessity to construct another structure with the changed parameters. The dynamic shift of EPs can be realized, for example, with nonlinearity \cite{Ramezanpour2021, Bai2026} or phase-change materials \cite{Dong2026}, but such approaches may strongly complicate system's dynamics. Recently, we have proposed to use linear anisotropic media to realize the EP tunability. In particular, we have analyzed the $\mathcal{PT}$-symmetric layered structures with a liquid-crystal layer and shown that its scattering-matrix EPs can be shifted by varying the optical-axis orientation with the external electric field \cite{Ramanovich2024}. Moreover, since anisotropy drives the polarization dependence of optical response, the resulting EPs provide a nontrivial possibility to manipulate light polarization, including two-sided absence of reflection and preferential transmission of a certain polarization at the so-called double EPs \cite{Ramanovich2025}, and to obtain the polarization-dependent mode exchange and $\pi$ phase jumps \cite{Ramanovich2026}.

An interesting class of anisotropic media are the magneto-optic ones with their optical properties controlled with the external magnetic field. Optical resonances can be used to enhance the magneto-optic effects as reported for many plasmonic and Mie-resonant nanophotonic structures \cite{Naghizadeh2026}. However, there is only a couple of relevant works using the magneto-optic effect to control non-Hermitian system's response. In Ref. \cite{Zhao2025}, the topological evolution of EPs in the band structure of two-dimensional magneto-optic photonic crystals is demonstrated by applying the external magnetic field. In Ref. \cite{Ruan2025}, the EP-based magnetic-field sensor is proposed with the magneto-optic medium providing the coupling between differently-polarized lossy modes of the Fabry-Perot cavity. The further examples of magnetically controlled non-Hermitian photonic structures include the topological transport in gyromagnetic photonic crystals \cite{Wang2025b} and the non-Hermitian skin effect in the two-dimensional photonic lattices \cite{Longhi2025}. There is also an entire subfield of non-Hermitian and EP-related effects in diverse magnonic and spintronic systems containing optical or optomechanical components \cite{Zhang2019b, Qian2024, Kim2025, Yu2025, Wang2026}. As far as we know, the magneto-optic effect was not applied to control the scattering-matrix EPs in layered non-Hermitian systems yet.

In this paper, we consider the archetypal three-layer $\mathcal{PT}$-symmetric photonic structure containing a magneto-optic material placed between the balanced loss and gain layers. Using theoretical analysis of the scattering matrix behavior, we show that the magneto-optic effect can be used to shift EPs providing the lines of EPs (exceptional lines) in the parameter space. This effect can serve as a basis for the tunable magnetic-field perturbation sensing with the level of magnetic field adjusted by the angle of light incidence. We also discuss the influence of magnetic field on the number of EPs and the mode-exchange effect in our layered structure. Finally, we discuss the possible realizations and extensions of the idea of magnetically-controlled EPs.

\section{Structure description}

\begin{figure}[t!]
    \centering
    \includegraphics[width=8 cm]{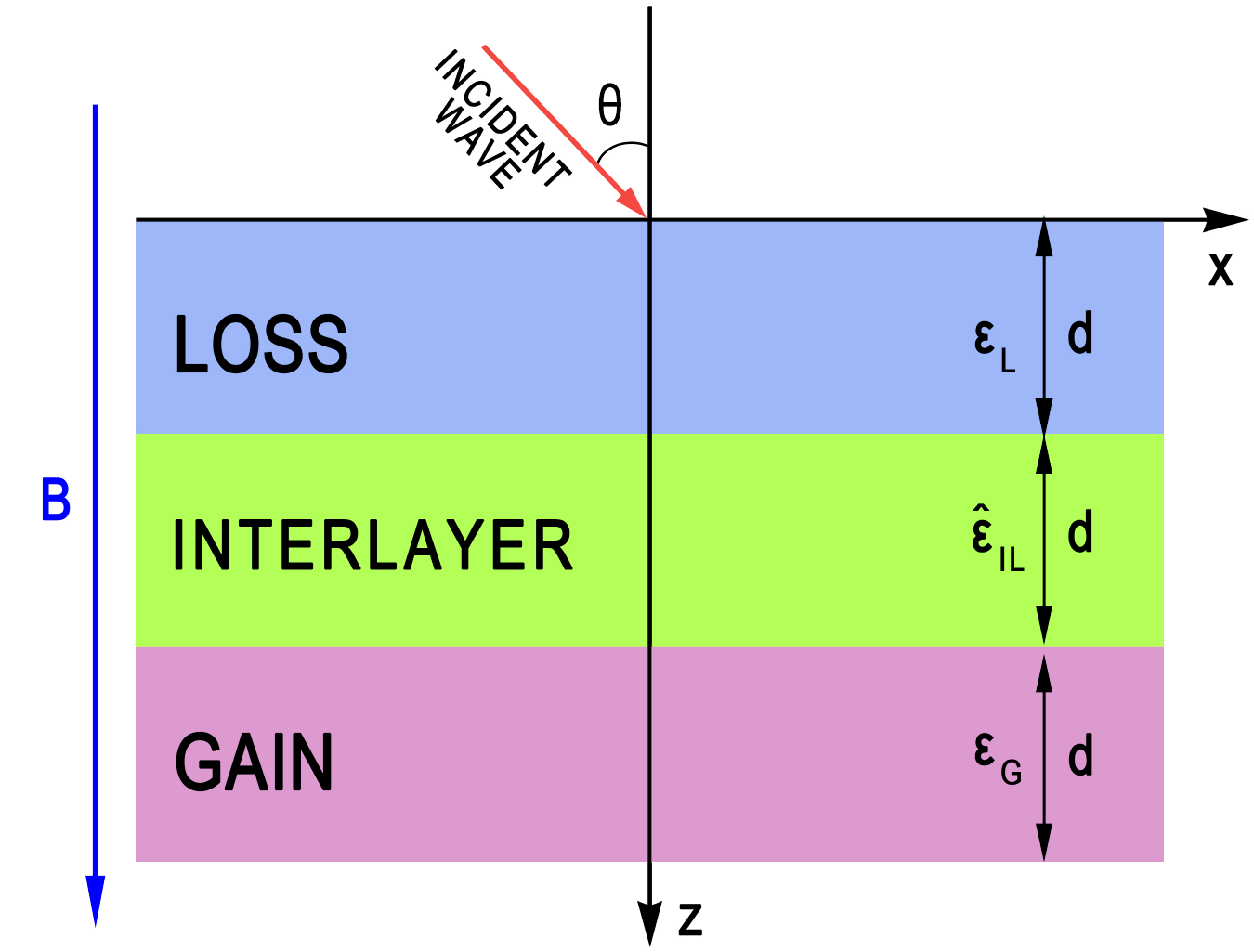}
    \caption{$\mathcal{PT}$-symmetric trilayer structure with the magneto-optic interlayer sandwiched between the layers with loss and gain. The magnetic field is oriented along the $z$ axis, orthogonal to the layers plane.}
    \label{fig1}
\end{figure}

The structure we study in this work is the $\mathcal{PT}$-symmetric trilayer with the isotropic outer layers and the magneto-optic material sandwiched between them (see Fig. \ref{fig1}). The outer layers are characterized by the permittivity $\varepsilon_{L/G}=(n \pm i \kappa)^2$ with the loss (gain) value governed by the non-Hermiticity parameter $\kappa$ and the refractive index $n$ ($n=2$ in our calculations). The thickness of all the layers is the same, $d=3$ $\mu$m, if otherwise is not stated.

Under a static magnetic field $B$ applied along the propagation direction $z$, the permittivity tensor of the magneto-optic interlayer is given as follows: 
\begin{equation}
\hat \varepsilon_{IL} = \left( \begin{array}{cccc} \varepsilon_{M} & i g & 0  \\ -i g & \varepsilon_{M} & 0 \\ 0 & 0 & \varepsilon_{M}   \end{array} \right), \label{DielectricPermMO}
\end{equation}
where $\varepsilon_M$ is the diagonal term and $g$ is the gyration vector’s absolute value serving as the magneto-optic coefficient. To be concrete, we adopt in our calculations the value $\varepsilon_M=4.75$ corresponding to the permittivity of the magnetic dielectric Bi-substituted yttrium iron garnet (Bi:YIG) \cite{Belotelov2005} at the wavelength $\lambda=1.55$ $\mu$m. We neglect the loss in the magneto-optic material assuming $\varepsilon_M$ to be real; the influence of loss is discussed further in Section \ref{Disc}. The non-diagonal components of the tensor \eqref{DielectricPermMO} depend on the magnetization $M$ of the material and change proportionally to the applied external magnetic field, $g \sim B$. We suppose that the magnetic field is weak enough, so that the effects of saturation are not observed.

We emphasize the difference of this case from our previous considerations of the trilayer structures with the liquid-crystal middle layer. In particular, the permittivity tensor of that layer can be written as
%
\begin{eqnarray}
&&\hat \varepsilon_{IL} =\varepsilon_o \left( \begin{array}{cccc} 1 & 0 & 0  \\ 0 & 1 & 0 \\ 0 & 0 & 1 \end{array} \right) \\
&+& (\varepsilon_e-\varepsilon_o) \left( \begin{array}{cccc} 0 & 0 & 0  \\ 0 & \cos^2(\gamma) & -\cos(\gamma) \sin(\gamma) \\ 0 & -\cos(\gamma) \sin(\gamma) & \sin^2(\gamma)   \end{array} \right), \nonumber \label{DielectricPermE7}
\end{eqnarray}
where $\varepsilon_o$ and $\varepsilon_e$ are the permittivities for the ordinary and extraordinary waves, respectively, and $\gamma$ is the angle between the optical axis and the layers plane. We have shown previously that the position of EPs can be controlled with that latter angle \cite{Ramanovich2025}. In contrast to the identical real $\gamma$-controlled non-diagonal elements in the tensor \eqref{DielectricPermE7}, the non-diagonal terms in the permittivity tensor for the magneto-optic material \eqref{DielectricPermMO} are purely imaginary and have the opposing signs. This provides the dramatic distinction between the two cases.

To study light propagation inside the multilayered structures, we use the operator transfer-matrix method described in Refs. \cite{Ramanovich2024, Ramanovich2026}. This method is based on the evolution-operator formalism and allows us to determine the output fields from any input field taking into account polarization changes in the anisotropic media. The resulting transmittances $t$ and reflectances $r$ of the non-Hermitian anisotropic structure generally depend on the polarization and propagation direction of light. These values are then substituted into the scattering matrix as follows,
\begin{equation}
\hat S = \left( \begin{array}{cccc} t_{xx} & t_{xy} & r_{xx}^R & r_{xy}^R \\ t_{yx} & t_{yy} & r_{yx}^R & r_{yy}^R \\ r_{xx}^L & r_{xy}^L & t_{xx} & t_{xy}\\ r_{yx}^L & r_{yy}^L & t_{yx} & t_{yy}  \end{array} \right), \label{ScatMatAniso}
\end{equation}
where the subscripts denote the polarizations of input and output radiation (either co- or cross-polarized), and the superscripts are for the propagation direction of input light (incident from right or left) that influences reflectance, but not transmittance. This choice of the scattering matrix implies that there are generally $4$ eigenvalues and eigenvectors, which pairwise coalescence produces a double set of EPs as shown in detail in our previous works \cite{Ramanovich2024, Ramanovich2025}.

\begin{figure*}[t!]
    \centering
    \includegraphics[width=8.5 cm]{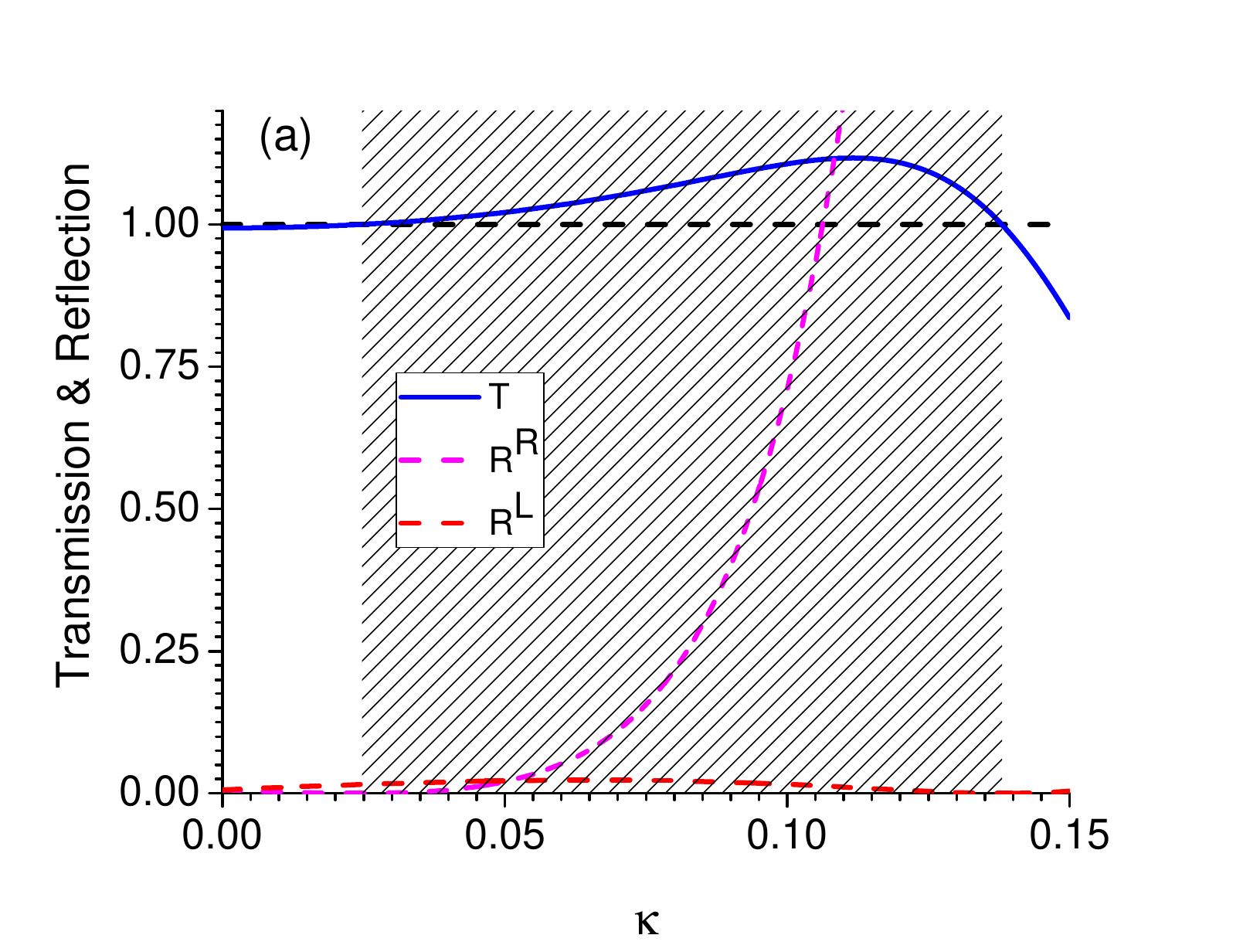}
    \includegraphics[width=8.5 cm]{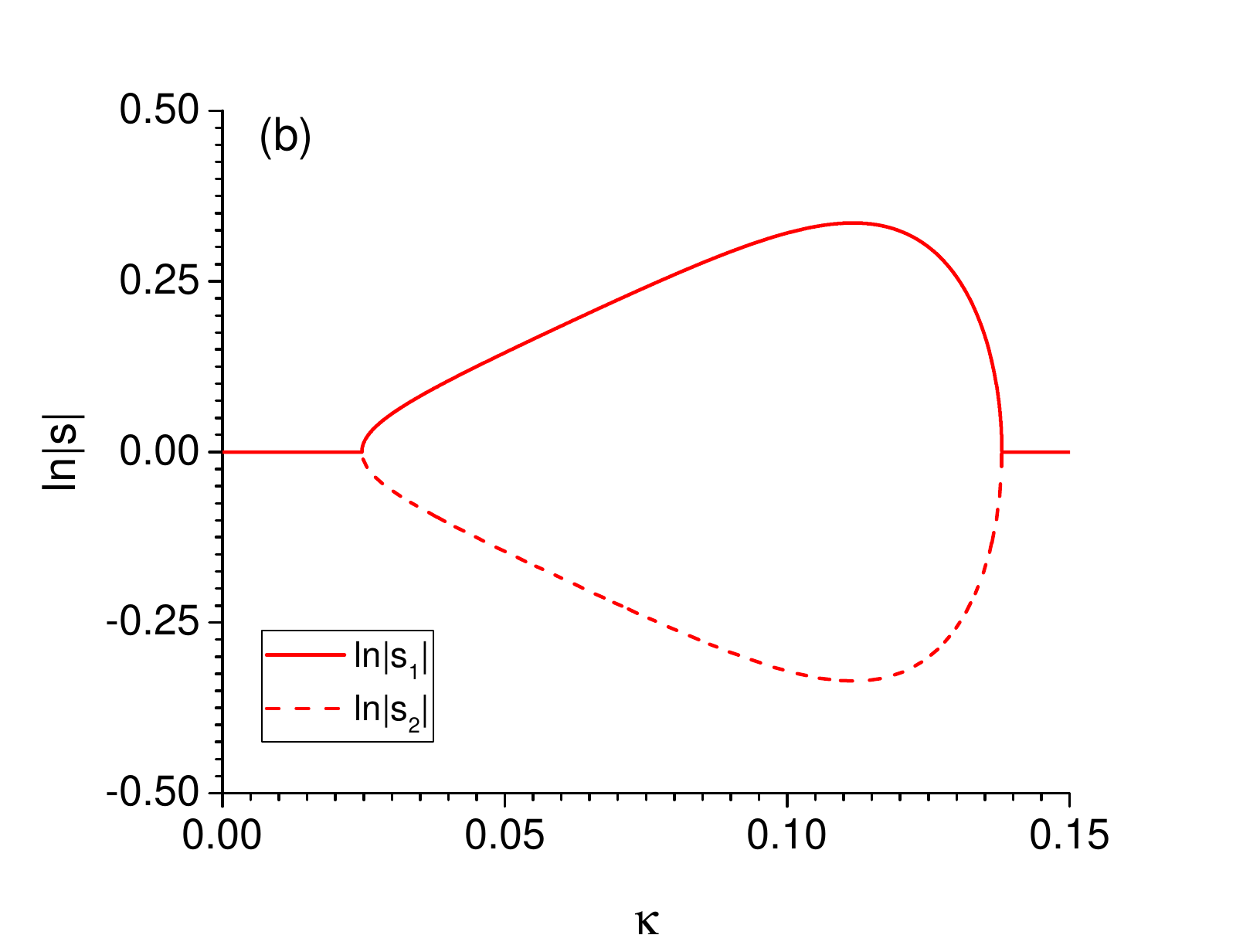}
    \includegraphics[width=8.5 cm]{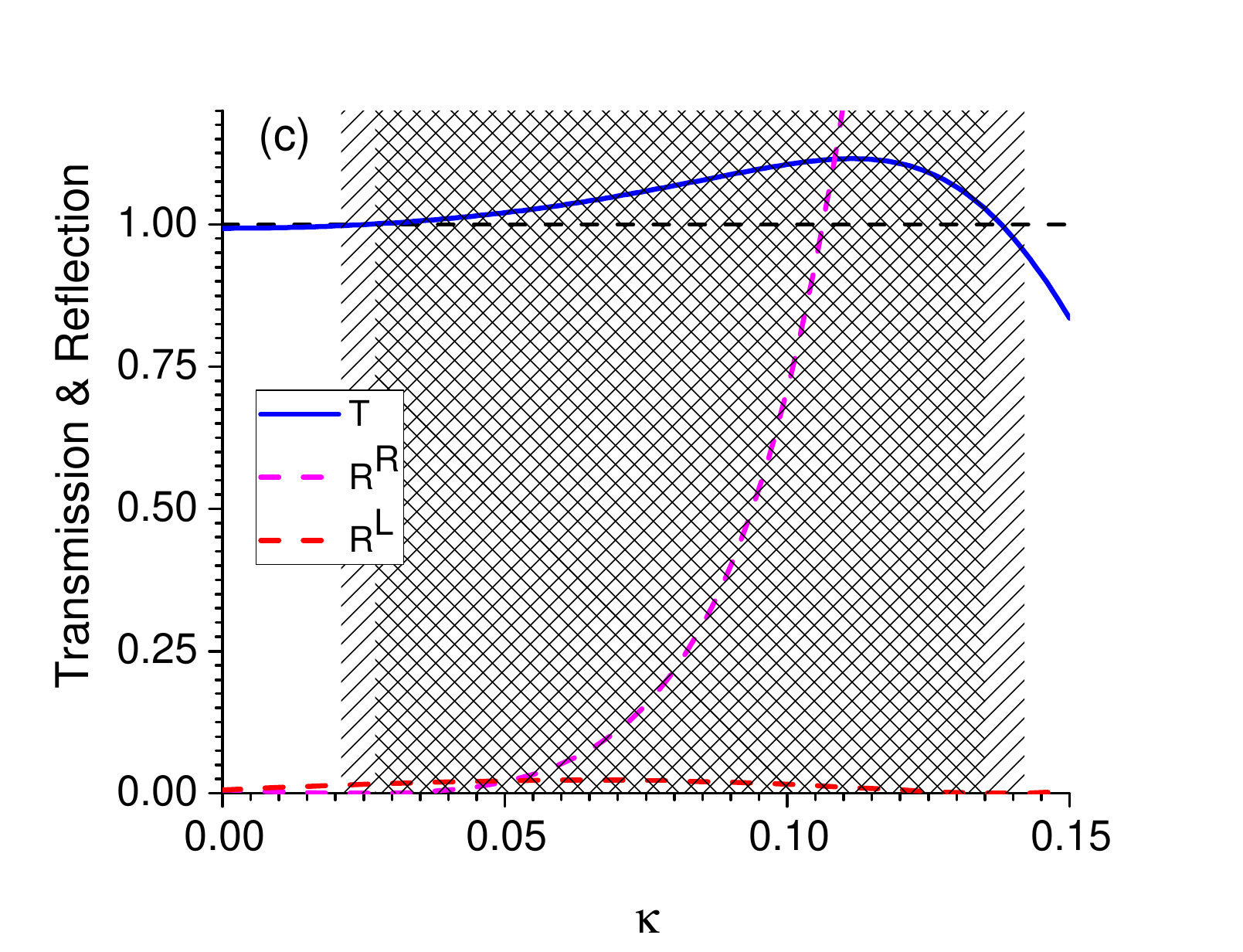}
    \includegraphics[width=8.5 cm]{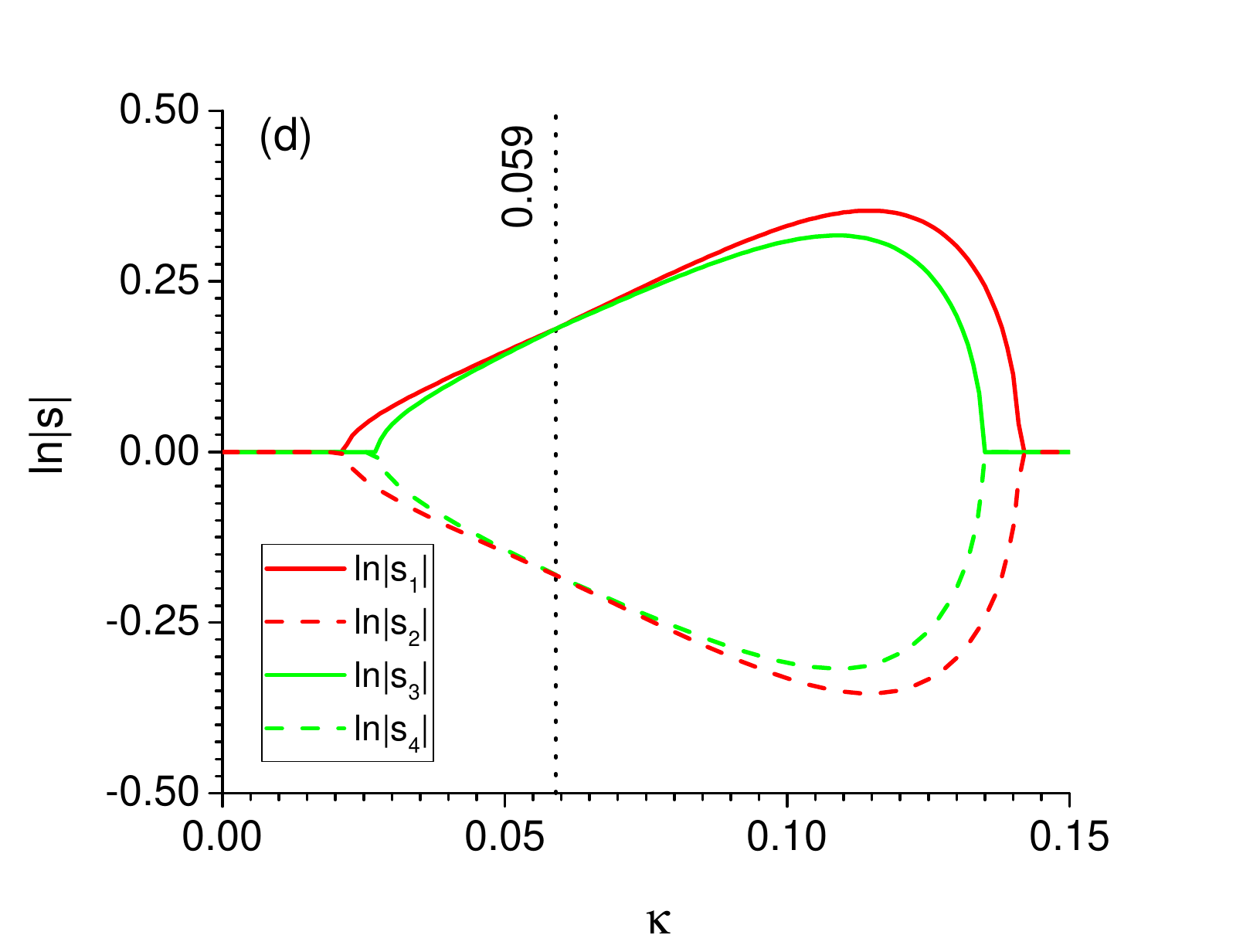}
    \caption{Transmission and reflection spectra for (a) the isotropic structure ($g=0$) and (c) the anisotropic structure ($g=0.005$) as a function of the non-Hermiticity parameter $\kappa$. (b) and (d) Logarithm of the scattering-matrix eigenvalues magnitude as a function of $\kappa$ for the isotropic and anisotropic structures, respectively. Shaded areas in panels (a) and (c) correspond to the $\mathcal{PT}$-symmetry-broken regions with $\ln|s| \neq 0$ in the panels (b) and (d). The vertical dotted line in the panel (d) marks the mode-exchange point.} 
    \label{fig2}
\end{figure*}

\section{EP shift with the magnetic field}

Let us start with the case of zero external magnetic field ($g=0$), so that the permittivity tensor \eqref{DielectricPermMO} takes on a diagonal form and the interlayer medium becomes isotropic. Figure \ref{fig2}(a) shows the polarization-independent transmission and reflection of the isotropic trilayer structure as a function of the non-Hermiticity parameter $\kappa$ for normally-incident radiation. One can see that there is a region of $\kappa$ values, where transmission is above unity. This region highlighted with hatching in Fig. \ref{fig2}(a) is limited on both sides by EPs and corresponds to the phase of broken $\mathcal{PT}$ symmetry as evidenced in Fig. \ref{fig2}(b): the scattering-matrix eigenvalues have unitary magnitude ($|s|=1$) in the $\mathcal{PT}$-symmetric phase and are inverse otherwise ($|s_1|=1/|s_2|$). The EPs, where unimodularity of the eigenvalues gets broken, correspond to the unity transmission ($T=|t|^2=1$) and single-sided reflectionlessness (either $R^{R}=|r^R|^2=0$, or $R^{L}=|r^L|^2=0$).

\begin{figure*}[t!]
    \centering
    \includegraphics[width=8 cm]{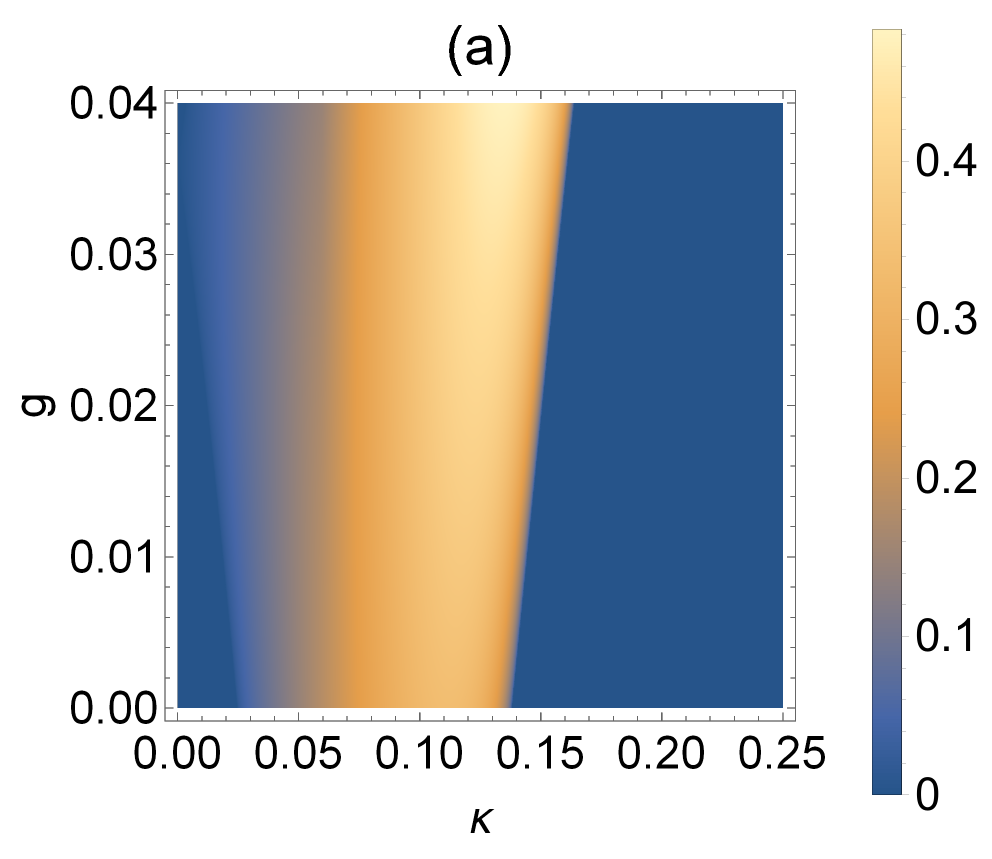}
    \includegraphics[width=8 cm]{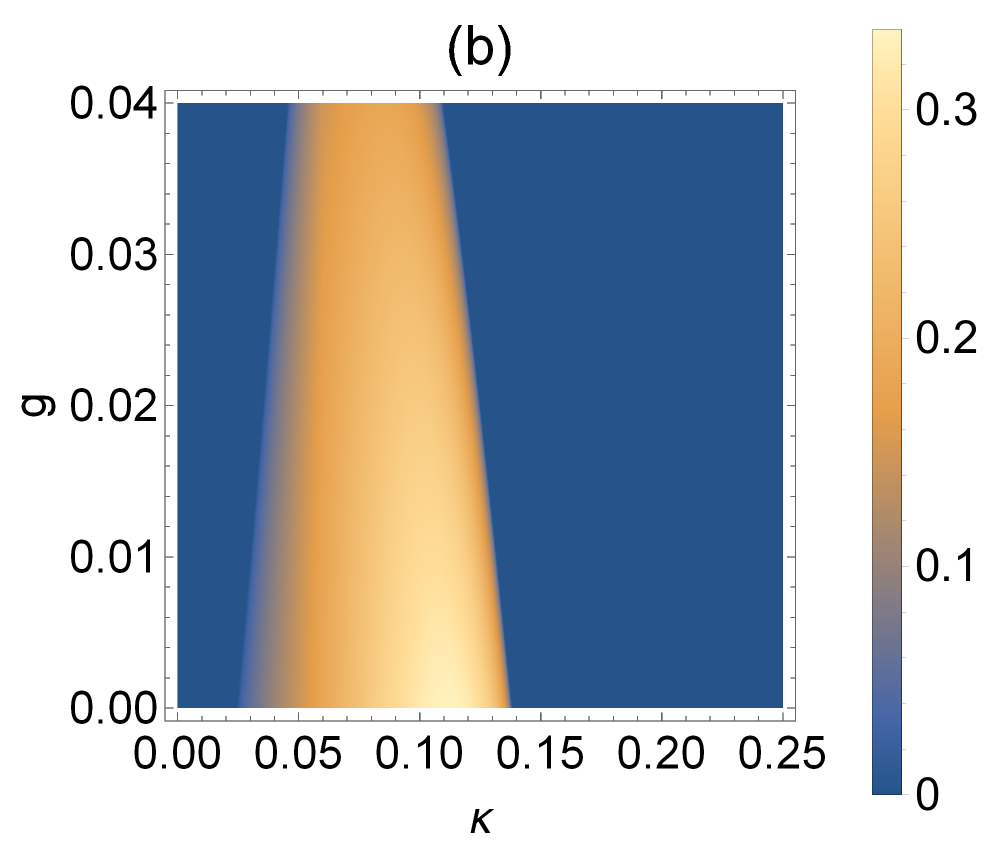}
    \includegraphics[width=8 cm]{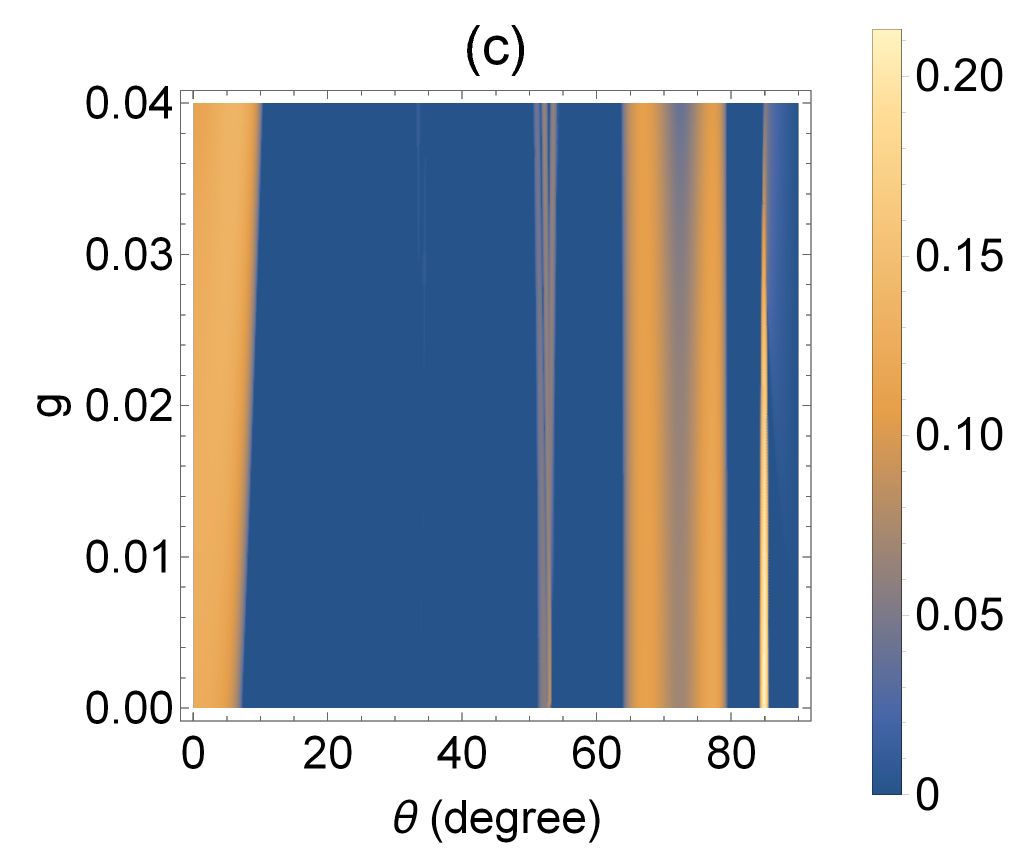}
    \includegraphics[width=8 cm]{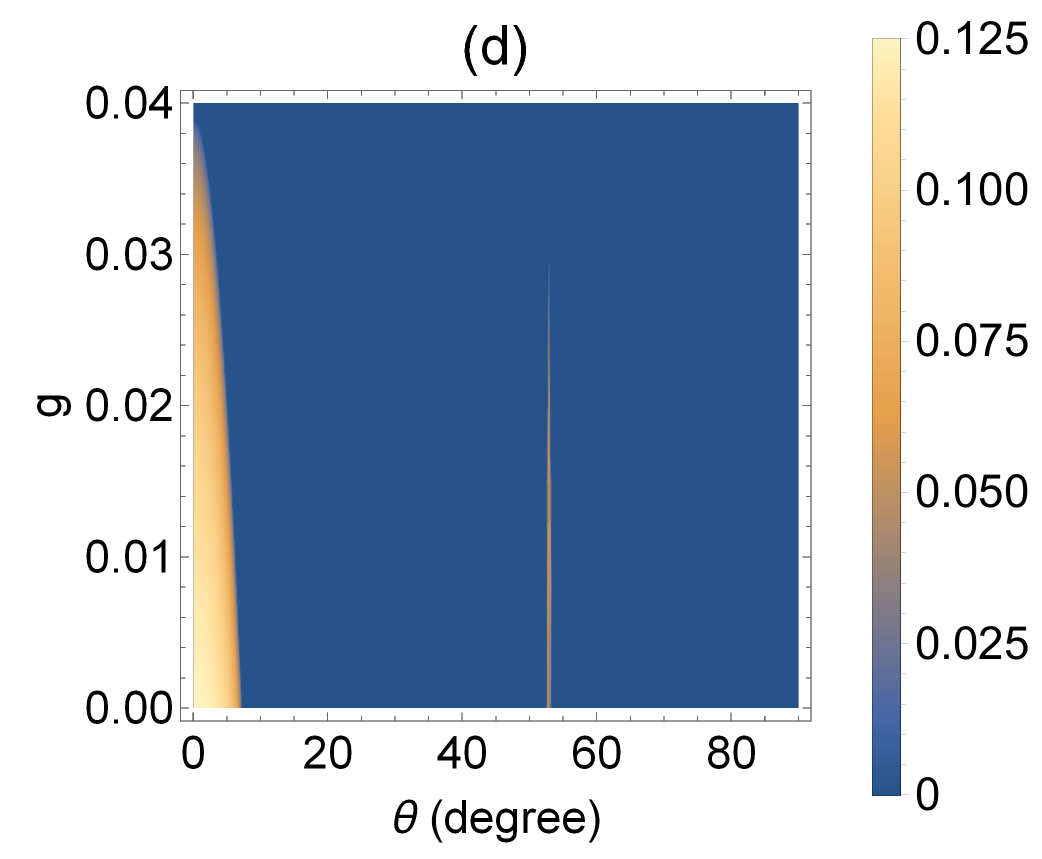}
    \caption{Density plot of the absolute values of the eigenvalue magnitude logarithm (a), (b) as a function of non-Hermiticity parameter $\kappa$ and magneto-optic coefficient $g$ for the incidence angle $\theta=0^{\circ}$; (c), (d) as a function of $\theta$ and $g$ for $\kappa=0.045$. Calculations are performed for (a), (c) the first and (b), (d) the second pair of eigenvalues.}
    \label{fig3}
\end{figure*}

Application of the magnetic field ($g \neq 0$) generates anisotropy in the interlayer medium and makes the response polarization-dependent. The resulting picture of transmission and reflection for $g=0.005$ as a function of the non-Hermiticity parameter $\kappa$ is shown in Fig. \ref{fig2}(c). Only the co-polarized transmissions and reflections ($T=T_{xx}=T_{yy}$, etc.) are shown here, since the cross-polarized values are much lower at this comparatively small anisotropy. The dependencies are very similar to those in the isotropic case (Fig. \ref{fig2}(a)). The small deviations, however, show the important changes of response due to anisotropy as can be seen from the comparison with the corresponding behavior of eigenvalues presented in Fig. \ref{fig2}(d). For the anisotropic case, we have $4$ eigenvalues due to the $4 \times 4$ dimensionality of the scattering matrix and, hence, the number of EPs also doubles. As a result, we have two overlapping regions of broken $\mathcal{PT}$ symmetry shown as the shaded areas in Fig. \ref{fig2}(c). One can easily see that the EPs are already not associated with the unity transmission and zero reflection due to the nonzero nondiagonal elements of the scattering matrix. In addition, the so-called point of mode exchange \cite{Ramanovich2026}, where the trajectories of different eigenvalues touch, is also shown with the vertical dotted line in Fig. \ref{fig2}(d); we discuss such points in more detail in Section \ref{Exchange}.

The full picture of the $\mathcal{PT}$-symmetric regions (dark blue areas) and the regions of broken $\mathcal{PT}$ symmetry (bright yellow spots) on the plane $(\kappa, g)$ is shown in Figs. \ref{fig3}(a) and \ref{fig3}(b). The case of normal incidence is studied in these phase diagrams obtained for the two pairs of scattering-matrix eigenvalues. It is clearly seen that, for the certain values of $\kappa$, $\mathcal{PT}$ symmetry cannot be broken for any realistic values of the magneto-optic coefficient $g$. On the other hand, sometimes it remains broken for any magnetic field strength, even in the absence of external magnetic field ($g=0$). We are mostly interested in the boundary between the regions of preserved and broken symmetry -- the line formed by EPs (exceptional line). These lines are slightly inclined on the $(\kappa, g)$ plane, so that one can fix the non-Hermiticity parameter (e.g., at $\kappa=0.15$) and realize $\mathcal{PT}$-symmetry breaking by changing exclusively the magneto-optic parameter $g$.  

Another possibility is to fix loss and gain and vary the incidence angle $\theta$. The resulting phase diagrams in the plane $(\theta, g)$ for the the non-Hermiticity parameter $\kappa=0.045$ are shown in Figs. \ref{fig3}(c) and \ref{fig3}(d). On the large scale, the $\mathcal{PT}$-symmetry-broken regions (bright spots) seem as almost vertical stripes. Nevertheless, the exceptional lines separating the phases are slightly inclined, so that, in a certain range of angles, one can break $\mathcal{PT}$ symmetry simply by varying the magnetic field. The change in the position of EPs with increasing $g$ can be traced in the GIF-animation 1 in the Supplementary Material. The details of the exceptional lines behavior seen on the small scale and their meaning are discussed further.

\begin{figure*}[t!]
    \centering
    \includegraphics[width=8 cm]{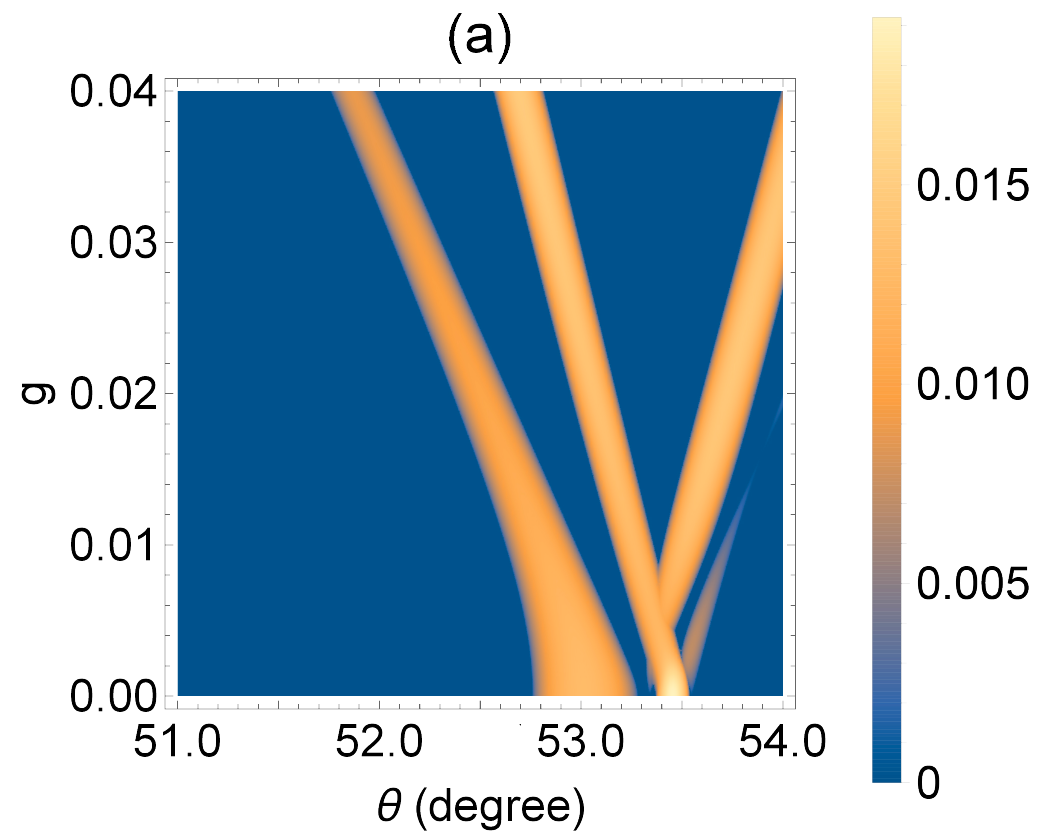}
    \includegraphics[width=8 cm]{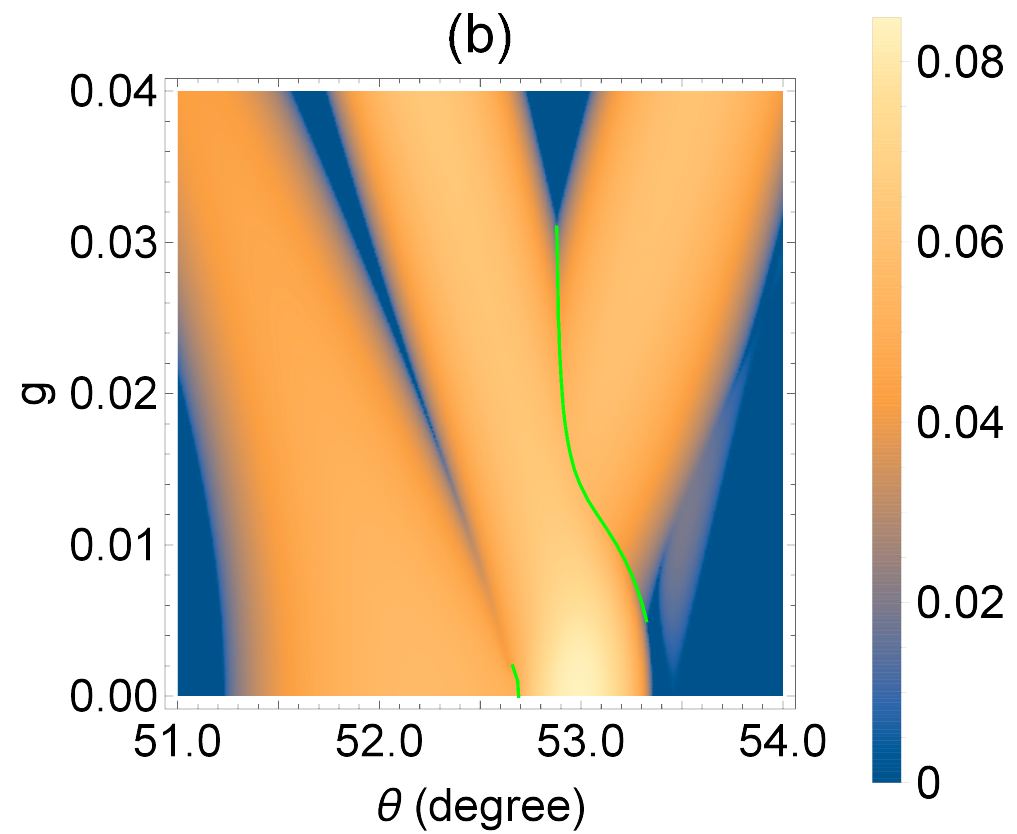}
    \caption{Density plot of the absolute values of the eigenvalue magnitude logarithm as a function of the incidence angle $\theta$ and the magneto-optic coefficient $g$ for the non-Hermiticity parameter of (a) $\kappa=0.015$ and (b) $\kappa=0.045$. Calculations are performed for the first pair of eigenvalues. Green lines in the panel (b) show the mode-exchange lines.}
    \label{fig4}
\end{figure*}

\section{Changing the number of EPs with the magnetic field}

We have seen in the previous section that EPs of our non-Hermitian structure containing magneto-optic material can be shifted with the magnetic field. As a further question, one can ask: is it possible to change the number of EPs via the external magnetic field in the structure with the fixed geometry and non-Hermiticity? Such a possibility would be valuable, since it is not an easy task to tune loss and gain after the fabrication of a structure. To answer this question, we take a closer look at the phase diagrams at the fixed non-Hermiticity parameter $\kappa=0.045$ shown in Figs. \ref{fig3}(c) and \ref{fig3} (d). The part of the regions of broken $\mathcal{PT}$ symmetry limited by the incidence angles $51^\circ \leq \theta \leq 54^\circ$ is depicted in Fig. \ref{fig4}(b). It is obvious that these regions of broken symmetry have rather complicated structure. In particular, one can see not only the shift in the EP position, but also the change in the number of EPs as we gradually increase the coefficient $g$ (see also the GIF-animation 1 in the Supplementary Material). Indeed, the splitting of the regions of broken $\mathcal{PT}$ symmetry is clearly observed with growing $g$, so that in this narrow range of $\theta$ the number of EPs separating different phases increases from $2$ (in the magnetic-free case, $g=0$) to $6$ (at $g=0.04$).

In Figs. \ref{fig4}(a) and \ref{fig4}(b), we present the phase diagrams for the structures with different values of loss and gain $\kappa$. It is seen that decreasing $\kappa$ results in the more narrow and slightly shifted regions of broken $\mathcal{PT}$ symmetry. However, the general picture of splitting regions and increasing the number of EPs remains intact. This means that the requirements for the non-Hermiticity strength can be strongly relaxed without qualitative change in system's behavior. This observation is important for possible experimental testing of the reported results.

\section{The principle of magnetic-field perturbation sensing}

\begin{figure*}[t!]
    \centering
    \includegraphics[width=8 cm]{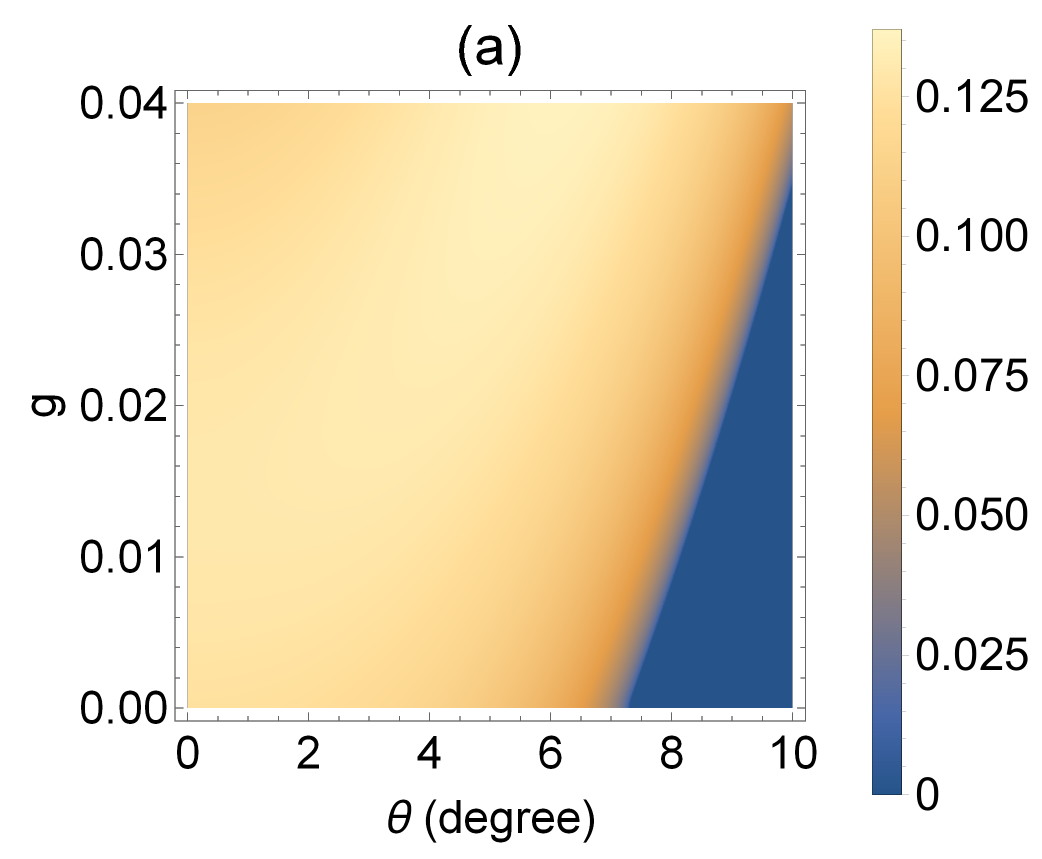}
    \includegraphics[width=8 cm]{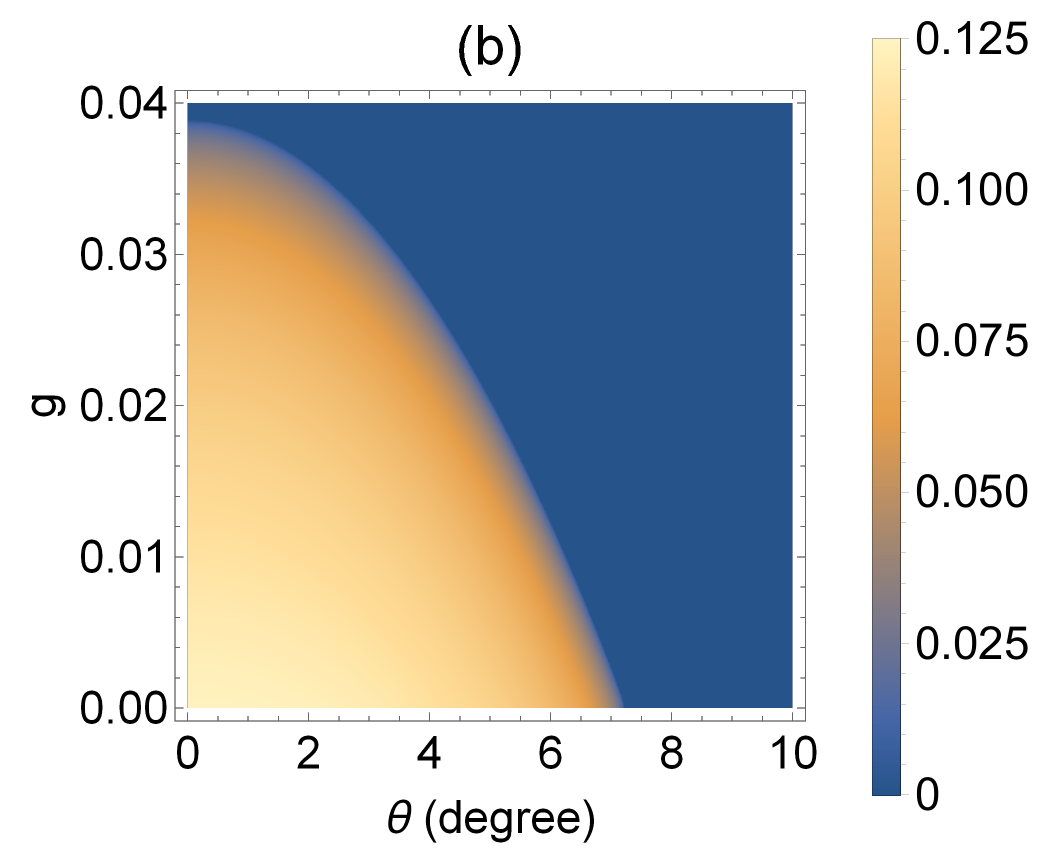}
    \caption{Density plot of the absolute values of the eigenvalue magnitude logarithm as a function of the incidence angle $\theta$ and the magneto-optic coefficient $g$ with $\kappa=0.045$. Calculations are performed for (a) the first and (b) the second pair of eigenvalues.}
    \label{fig5}
\end{figure*}

\begin{figure*}[t!]
    \centering
    \includegraphics[width=8.5 cm]{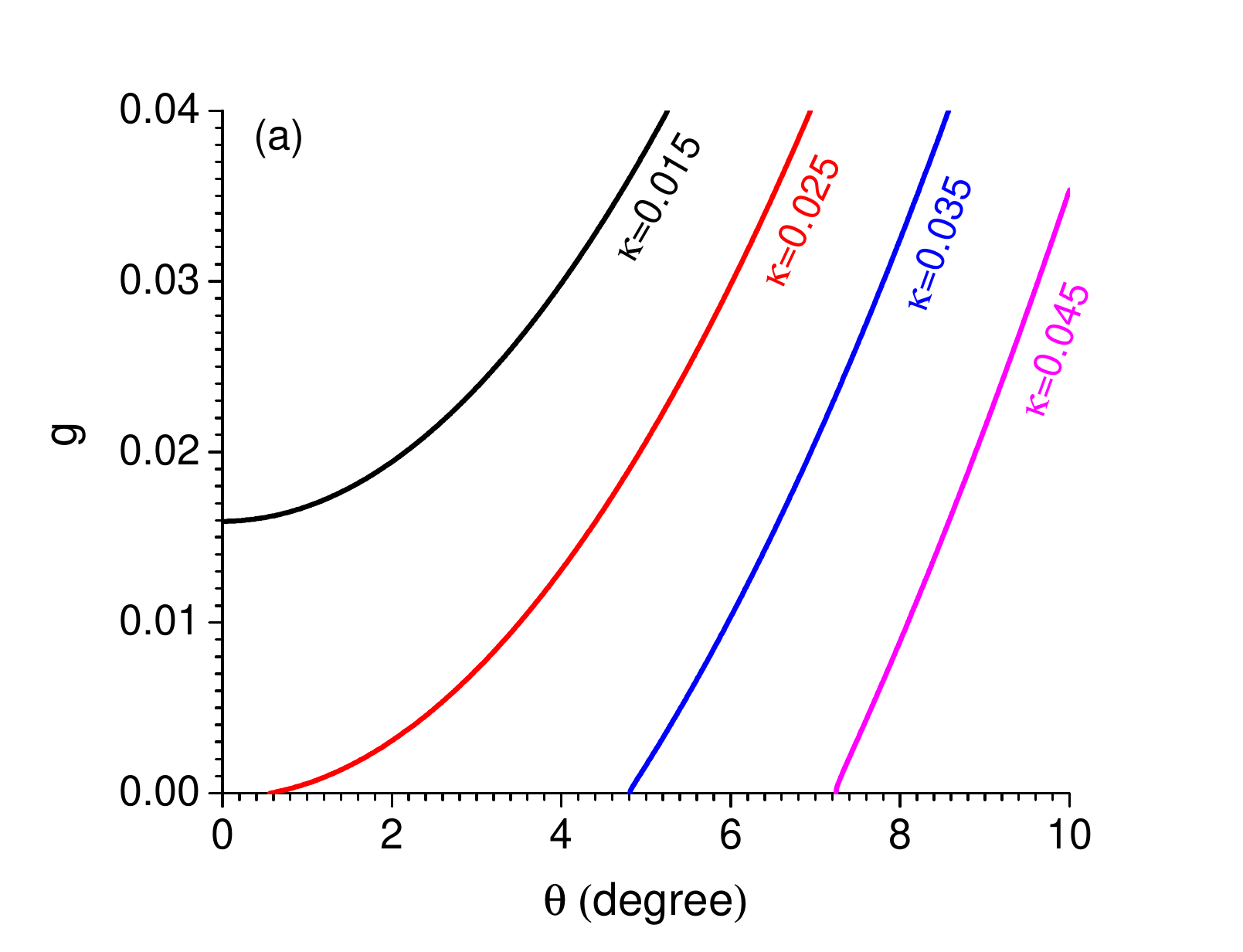}
    \includegraphics[width=8.5 cm]{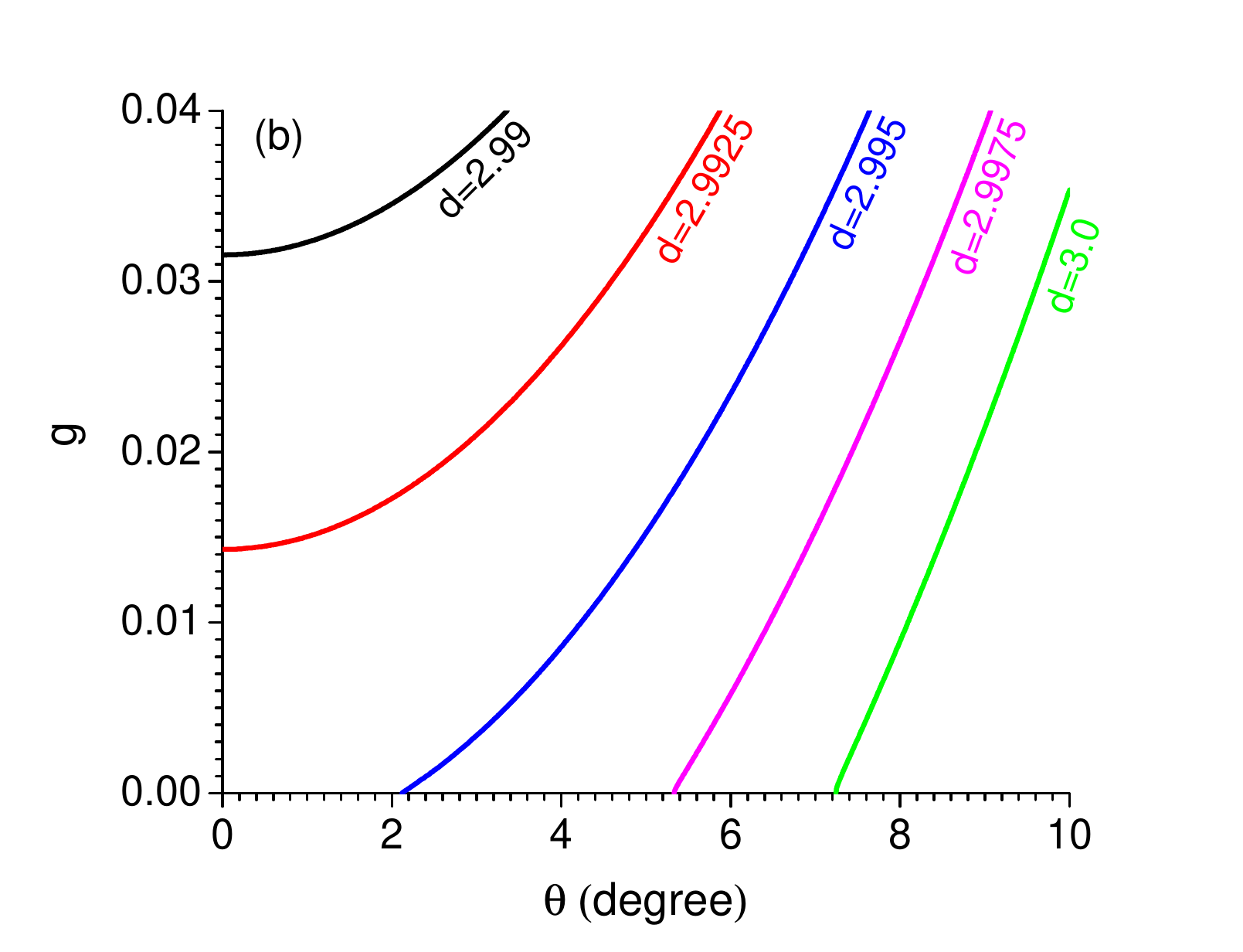}
    \caption{Exceptional lines for the first pair of eigenvalues as a function of the incidence angle $\theta$ and the magneto-optic coefficient $g$ for the structures with different values of (a) the non-Hermiticity parameter $\kappa$ and (b) the thickness of all layers $d=d_{L/G}=d_{IL}$ (in $\mu$m).}
    \label{fig6}
\end{figure*}

As we have noted analyzing the phase diagrams in Figs. \ref{fig3} and \ref{fig4}, the exceptional lines separating the $\mathcal{PT}$-symmetric and $\mathcal{PT}$-symmetry-broken phases are not strictly vertical, but are slightly inclined. This small inclination is of dramatic importance for the possibility to reach an EP by changing exclusively the external magnetic field. This makes the considered structures perspective for magnetic-field sensing. Indeed, EPs are known to provide the enhanced sensitivity to an external perturbation due to sublinear splitting of eigenvalues. In our case, such a perturbation is provided by the magnetic field deviation from the specific level corresponding to the EP. It is in this sense that we write here about the magnetic-field perturbation sensing.

Moreover, the inclination of the exceptional line mentioned above means that one can rationally choose the desired level of magnetic field, in the vicinity of which the magnetic-field perturbations can be sensed with enhanced sensitivity. For example, by producing the trilayer structures with the different non-Hermiticity parameters $\kappa$, it is possible to realize EPs at different values of the magneto-optic coefficient $g$ (see Figs. \ref{fig3}(a) and \ref{fig3}(b)) and, hence, enhance sensitivity to the perturbations near the magnetic fields of different strengths. Let us consider in more detail another possibility -- the case of the fixed $\kappa=0.045$ and the variable incidence angle $\theta$. The partial phase diagrams (for the small $\theta$s) is shown in Fig. \ref{fig5}. One can see that the exceptional line for the first pair of eigenvalues (Fig. \ref{fig5}(a)) is almost a straight line, so that changing the incidence angle from approximately $7^\circ$ to $10^\circ$ results in the shift of the EP position from $g_{EP}=0$ to $0.04$. This means that the rational choice of $\theta$ allows us to realize enhanced sensing of the magnetic-field perturbation in a wide range of strengths, even very weak. The second pair of eigenvalues (Fig. \ref{fig5}(b)) gives the clearly nonlinear exceptional line, which is not so convenient for calibration. Nevertheless, it can be also used for sensing in the range of even smaller incidence angles ($\theta < 7^\circ$).

Furthermore, the form and position of exceptional lines can be tuned by the choice of non-Hermiticity or geometrical parameters of the structure as shown in Fig. \ref{fig6}. Indeed, one can see in panel (a) that decreasing $\kappa$ results in EPs at higher $g$ and lower $\theta$. The same is valid for the thinner thicknesses of layers as demonstrated in panel (b) (all the layers were changed simultaneously in this calculation, but similar results are observed for changing exclusively the middle-layer thickness; compare the the GIF-animations 2 and 3 in the Supplementary Material). The inclination of exceptional lines increases for thinner layers as well, so that one can more smoothly adjust the range of highest sensitivity with the monotonous change of incidence angle. On the other hand, an almost linear behavior of exceptional lines at $\kappa=0.045$ and $d=3$ $\mu$m is preferable from the practical point of view. In any case, the realistic sensing systems of this type should be calibrated first to take into account the actual behavior of the exceptional line.

\begin{figure}[t!]
    \centering
    \includegraphics[width=9 cm]{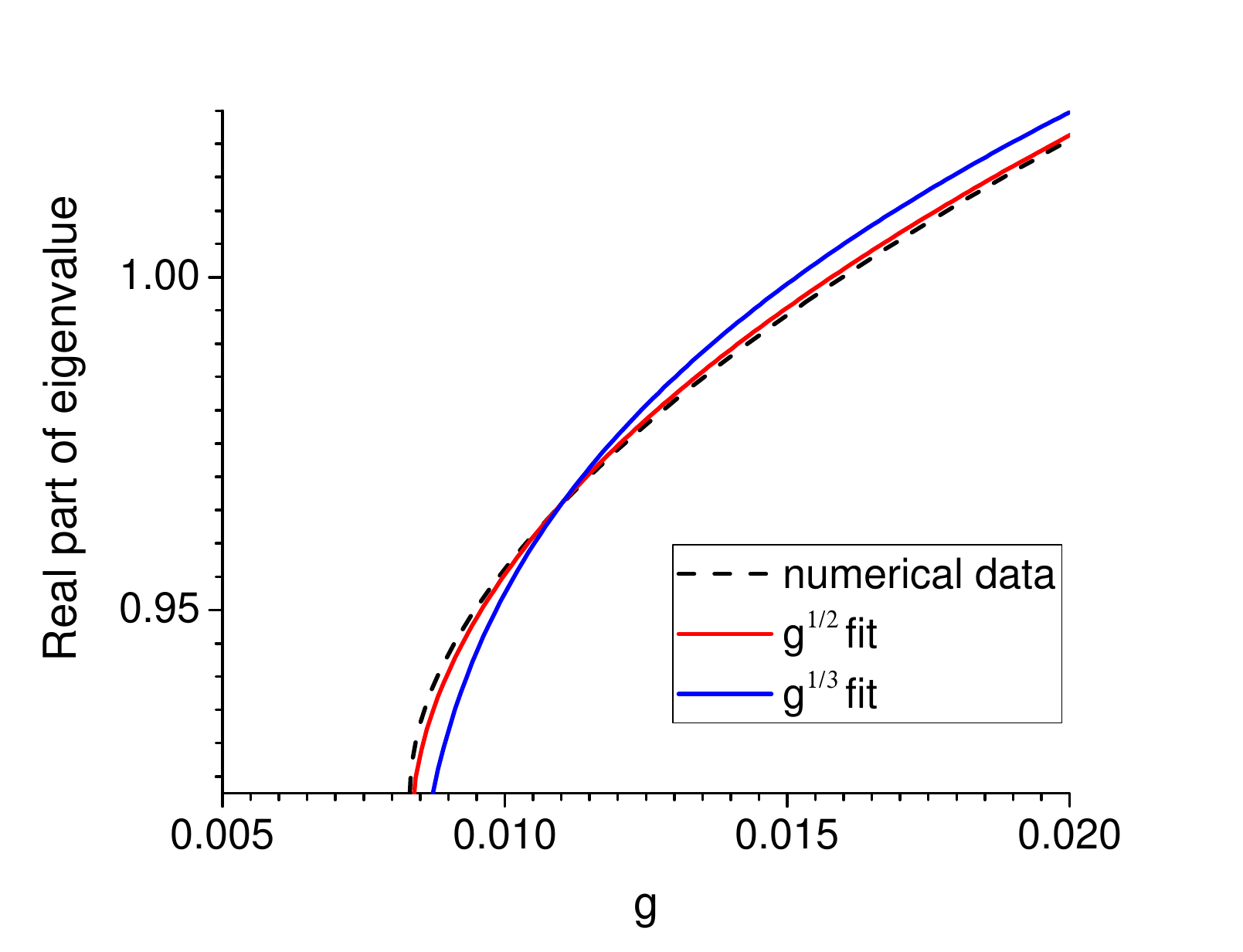}
    \caption{Behavior of the real part of the scattering-matrix eigenvalue in the vicinity of the EP seen in Fig. \ref{fig6}(a) at $\theta=8^\circ$ and $\kappa=0.045$. The fits with the square- and cube-root functions are shown as well.}
    \label{fig7}
\end{figure}

In order to prove that we have a sublinear (square-root) dependence on the magnetic field strength at EPs implying the enhanced sensitivity to perturbations, we plot in Fig. \ref{fig7} the splitting of eigenvalues as a function of the magneto-optic coefficient $g$. One can clearly see that the numerical data are perfectly fitted with the $\sqrt{g}$ dependence, whereas the $\sqrt[3]{g}$ fit has poor matching. Thus, in the vicinity of these $2$nd-order EPs, the higher sensitivity to magnetic field is possible in principle. Note that we do not deal here with the advanced topics of noise enhancing and signal-to-noise ratio, which should be taken into account in real devices.

\section{Mode-exchange lines \label{Exchange}}

We have mentioned previously the effect of mode exchange seen in Fig. \ref{fig2}(d). In this section, we discuss this effect in more detail with emphasis on the control with magnetic field. Let us recall that the number of eigenmodes and eigenvalues is doubled due to anisotropy, so that there are two regions of broken $\mathcal{PT}$ symmetry corresponding to each pair of eigenvalues. Mode exchange as an implicit manifestation of system's nontrivial topology appears at the point, where these regions touch, so that the curves for the different eigenvalues effectively change places (see the change of colors at the point marked with the vertical dotted line in Fig. \ref{fig2}(d)). We have shown earlier \cite{Ramanovich2026} that, at the mode-exchange point, the co-polarized transmissions become equal ($T_{xx}=T_{yy}$). Here, this fact is not enough to find a unique mode-exchange point, since the co-polarized transmissions are equal everywhere (see Fig. \ref{fig2}(c)) due to the specific (quasi-symmetric) form of the permittivity tensor \eqref{DielectricPermMO}. Nevertheless, the effect of mode exchange clearly takes place in our system.

It is of further interest to see how the mode-exchange points can be controlled with the magnetic field and, moreover, form the mode-exchange lines. Indeed, we see in Fig. \ref{fig4}(b) that increase in the magneto-optic coefficient $g$ results in splitting the region of broken $\mathcal{PT}$ symmetry. The green line marking this splitting consists of mode-exchange points. In other words, the $\mathcal{PT}$-symmetry-broken regions for the first and second pairs of eigenvalues are touching not in a separate point, but along a whole line, which can be called the mode-exchange line. We believe that the possibility for the formation of mode-exchange lines appears in our system due to the specific anisotropic response preserving the equality of co-polarized transmissions as mentioned above.

\section{Discussion \label{Disc}}

Although we do not consider the structure composed from the specific materials and do not predict any concrete magnetic fields perturbations of which could be sensed, it is instructive to discuss the realistic range of parameters that can be reached with our approach. The main magneto-optical characteristic is the non-diagonal element of permittivity tensor $g$, or the gyration vector’s absolute value. In the theoretical studies such as Ref. \cite{Zhao2025}, its magnitude is of the order of several hundredths (up to $0.1$) corresponding to the peak values reported for such materials as Bi:YIG \cite{Wettling1976}, Ce:YIG \cite{Onbasli2016}, and GdFe \cite{Jesenska2019}. Of course, the value of $g$ strongly depends on the wavelength, so that the specific parameters of the structure (such as layer thicknesses) should be carefully adjusted to obtain the expected response at the desired wavelength. A stronger magneto-optic effect is expected for the case of multiple magnetic layers \cite{Inoue1997} -- the situation worth to be studied in the future.

For the magnetic materials like mentioned above, the linear dependence of magneto-optic parameter on the magnetization (or external magnetic field) is usually implied \cite{Belotelov2005, Lyubchanskii2008}, $g \sim M/M_s \sim B/B_s$, where $M_s$ and $B_s$ are the saturation magnetization and field induction, respectively. The saturated state corresponds to all the magnetic domains oriented in the direction of field and is characterized by the maximal $g$ (this value is usually reported in the literature). Therefore, the saturation field is an important parameter governing the limits of linear behavior and, hence, the operational range of magnetic-field sensing. The typical values of saturation magnetization are of the order of $0.1 - 1$ T \cite{Handbook}. For example, for Ce:YIG \cite{Onbasli2016}, the saturation field of several kOe is reported, equivalent to several tenths of a T. 

\begin{figure}[t!]
    \centering
    \includegraphics[width=9 cm]{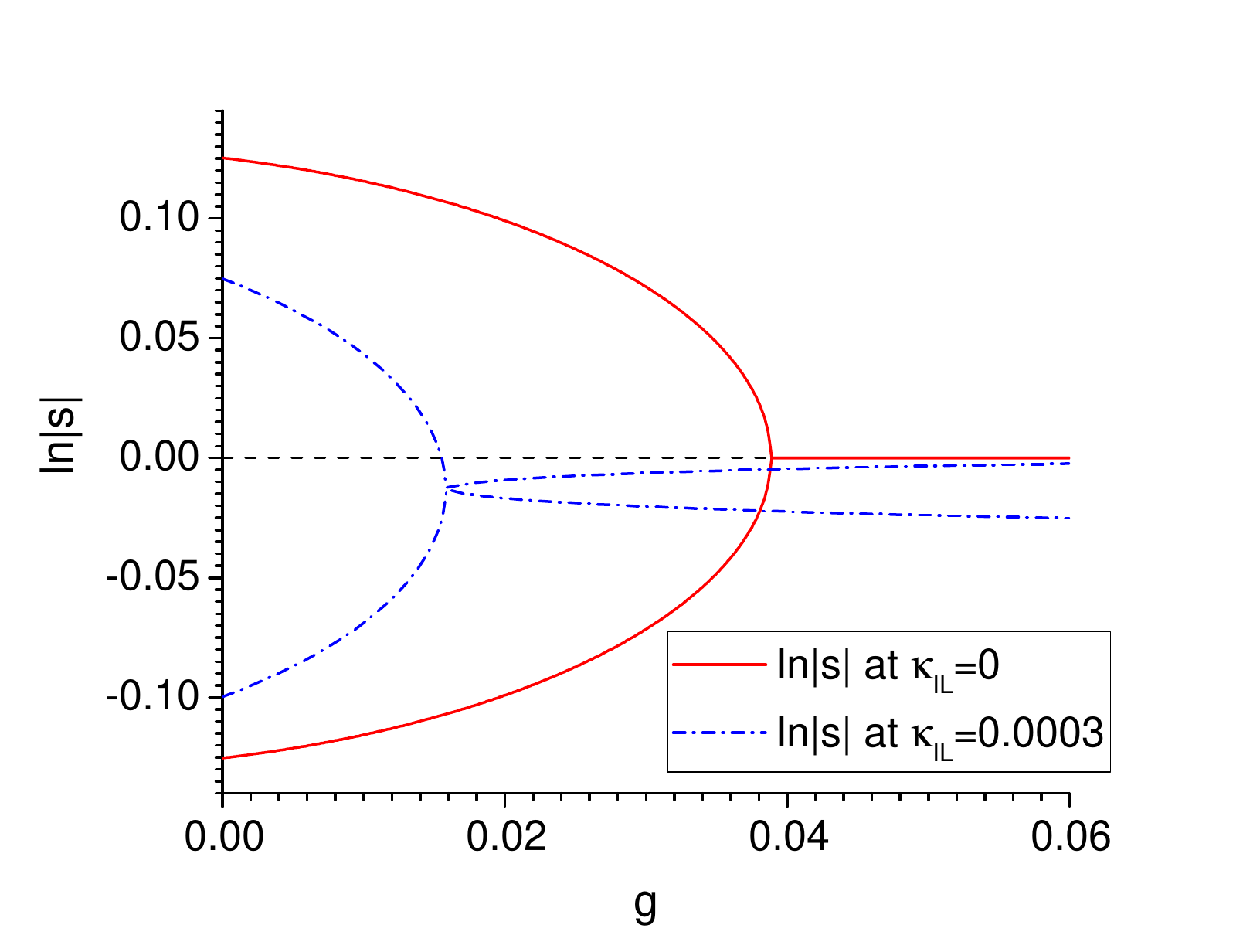}
    \caption{Logarithm of the scattering-matrix eigenvalues magnitude as a function of the magneto-optic coefficient $g$ for the structures with the lossless interlayer ($\kappa_{IL}=0$, red solid lines) and the lossy interlayer ($\kappa_{IL}=0.0003$, blue dash dotted lines). These two cases slightly differ in the thickness of gain layer, which is $d_{G}=3$ $\mu$m and $d_{G}=3.018$ $\mu$m, respectively.}
    \label{fig8}
\end{figure}

Next, we should discuss the influence of loss in the magneto-optical material on the reported effect of EP tuning. Magnetic dielectrics, such as Bi:YIG, have relatively low loss in comparison with ferromagnetics \cite{Naghizadeh2026}, so that we have focused on such materials above assuming them lossless. The loss can be taken into account by addition of an imaginary part to the diagonal elements of permittivity tensor \eqref{DielectricPermMO}, so that $\varepsilon_M=(n_M+i\kappa_M)^2$ with $n_M=\sqrt{4.75}$. Obviously, with such an addition, the system ceases to be $\mathcal{PT}$-symmetric and the regions of symmetric and symmetry-broken phases disappear. Nevertheless, EPs are still possible in such systems, since they are still non-Hermitian. As an example, we show in Fig. \ref{fig8} the behavior of scattering-matrix eigenvalues for the lossless ($\kappa_M=0$) and lossy ($\kappa_M=0.0003$) cases. In the lossless case, one can see the usual EP associated with $\mathcal{PT}$-symmetry breaking, where the scattering-matrix eigenvalues magnitude ceases to be unity (see the red solid lines). On the contrary, in the lossy case, the EP loses the meaning of a point of $\mathcal{PT}$-symmetry breaking, but, nevertheless, it can be restored (see blue dash dotted lines) due to the small increase of the gain-layer thickness from $d_{G}=3$ $\mu$m to $3.018$ $\mu$m to compensate for the loss in the middle layer. The resulting EP moves down from the zero line and can be reached at the lower values of magneto-optic coefficient than for the purely lossless interlayer (we have also directly checked that both real and imaginary parts of eigenvalues coalesce at the touching point seen in Fig. \ref{fig8}). In fact, the purely passive analogs of $\mathcal{PT}$-symmetric systems with loss and gain represented by the different levels of loss can be also used to realize EPs and control them with the magnetic field. This case, however, deserves a separate detailed study.

Finally, it is worth to emphasize that our calculations provide only the proof of principle for the magnetic-field perturbation sensing. We consider only the magnetic fields oriented along the $z$ axis, whereas in realistic situations it can be directed arbitrarily. A more detailed analysis with the permittivity tensor having other nondiagonal elements is needed to fully describe the general situation. Such a generalization can be used to lay the foundations for fully three-dimensional vectorial sensing of magnetic-field perturbations.

\section{Conclusion and outlook}

In conclusion, using the trilayer $\mathcal{PT}$-symmetric structure containing magneto-optic material as an example, we show that it is possible to tune the number and positions of EPs with the external magnetic field. Magneto-optic effect makes the material anisotropic and the response polarization-dependent. The number of EPs is also doubled due to increasing the dimensions of scattering matrix and generally depends on the anisotropy level proportional to the magnetic field. The position of EPs in the parameter space is weakly dependent on the magnetic field as well. Nevertheless, we show that even this weak dependence is enough for the original concept of magnetic-perturbation sensor with EP-driven enhancement of sensitivity. The distinctive feature of the proposed sensing scheme is the possibility of easy tuning the magnetic-field strength at the EP by changing the angle of incident waves. We expect that this approach will prove to be general and applicable to different types of non-Hermitian photonic systems.

Moreover, we expect that the proposed approach to tune EPs with the magnetic field as well as the principle of magnetic-field perturbation sensing based on this approach are quite general and can be applied to different types of non-Hermitian photonic systems. In particular, the systems described with the effective Hamiltonian (such as coupled waveguides or cavities) can be considered as a possible platform for magnetically tuned EPs. In contrast to the systems described with the scattering matrix as in this paper, the entirely passive Hamiltonian-based systems are easier to connect to the $\mathcal{PT}$-symmetric ones by simply extracting the common level of loss or gain from the Hamiltonian matrix. Of course, magnetically-driven EPs are possible in the non-Hermitian systems without $\mathcal{PT}$ symmetry as well.

\acknowledgements{The work was supported by the Belarusian State Program for Scientific Research “Photonics and Microelectronics” (Task 5.1.04).}

\end{document}